# GPS-based Vehicle Tracking System-on-Chip

Adnan I. Yaqzan†, Issam W. Damaj††, and Rached N. Zantout†††

†MSc, Computer and Communication Engineering Research and Development, Millenium Group Services, Beirut, Lebanon
 adnan5884@hotmail.com

†† PhD MEng BEng MIEEE Assistant Professor of Computer Engineering Program Lead Division of Sciences and Engineering American University of Kuwait  idamaj@auk.edu.kw Home: http://academics.idamaj.net

††† Ph.D. Associate Professor College of Computer and Information Sciences Prince Sultan University P. O. Box 66833, Riyadh, 11586 Kingdom of Saudi Arabia RZANTOUT@PSU.EDU.SA   WebPage: http://info.psu.edu.sa/psu/cis/rzantout

*Abstract*—Modern powerful reconfigurable systems are suited in the implementation of various data-stream, data-parallel, and other applications. An application that needs real-time, fast, and reliable data processing is the global positioning system (GPS)-based vehicle tracking system (VTS). In this paper, we build on a recently produced VTS (The Aram Locator) offering a system-on-chip (SOC) replacement of the current microcontroller-based implementation. The proposed SOC is built on a field programmable gate array (FPGA) promising a cheaper design, a more cohesive architecture, a faster processing time and an enhanced system interaction. Different designs, and their hardware implementations, are proposed with different levels of integration. Performance analysis and evaluation of the investigated designs are included.

## I. INTRODUCTION

After a great evolution, reconfigurable systems fill the flexibility, performance and power dissipation gap between the application specific systems implemented with hardwired Application Specific Integrated Circuits (*ASICs*) and systems based on standard programmable microprocessors. Reconfigurable systems enable extensive exploitation of computing resources. The reconfiguration of resources in different parallel topologies allows for a good matching with the inherent intrinsic parallelism of an algorithm or a specific operation. The reconfigurable systems are thus very well-suited in the implementation of various data-stream, data-parallel, and other applications. The introduction of a new paradigm in hardware design called Reconfigurable Computing *(RC)* offers to solve any problem by changing the hardware configurations to offer the performance of dedicated circuits. Reconfigurable computing enables mapping software into hardware with the ability to reconfigure its connections to reflect the software being run. The ability to completely reprogram the computer's hardware implies that this new architecture provides immense scope for emulating different computer architectures [1], [2], [3].

The progression of field programmable gate arrays *(FPGAs) RCs* has evolved to a point where *SOC* designs can be built on a single device. *FPGA* devices have made a significant move in terms of resources and performance.

The contemporary *FPGA*s have come to provide platform solutions that are easily customizable for system connectivity, digital signal processing *(DSP),* and data processing applications [4].

As the complexity of *FPGA*-based designs grow, a need for a more efficient and flexible design methodology is required. One of the modern tools (also used in the proposed research) is *Quartus II*, started with *Altera. Quartus II* is a compiler, simulator, analyzer and synthesizer with a great capability of verification and is chosen to be used for this implementation. It can build the verification file from the input/output specification done by the user. *Quartus II* design software provides a complete, multiplatform design environment that easily adapts to your specific design needs. It is a comprehensive environment for system-on-a-programmable-chip (*SOPC*) design [5], [6].

An application that needs real-time, fast, and reliable data processing is GPS-based vehicle tracking. In this paper, we build on a recently produced *VTS* (The Aram Locator) offering a *SOC* replacement of the microcontroller-based implementation. Although the microcontroller-based system has acceptable performance and cost, an *FPGA*-based system can promise less cost and a more cohesive architecture that would save processing time and speeds up system interaction.

This paper is organized so that Section 2 presents the currently available existing microprocessor-based *VTS* and its proposed update. Section 3 proposes different designs and implementations with different levels of integration. In Section 4, performance analysis and evaluation of results are presented. Section 5 concludes the paper by summarizing the achievements and suggesting future research.

## II. Upgrading the Aram Locator GPS System

One recently implemented *VTS* is the Aram Locator [5], [7]. It consists of two main parts, the Base Station *(BS)* and the Mobile Unit *(MU)*. The *BS* consists of a *PIC* Microcontroller based hardware connected to the serial port of a computer. The *MU* is a self-contained *PIC* Microcontroller based hardware and a *GPS* module. The latter would keep track of all the positions traversed by the vehicle and records them in its memory. The system has a great storage capacity, and could perform a significant recording with a considerable sampling rate. The mobile unit *(MU)* of the addressed *Aram Locator* consists of two communicating microcontrollers interfaced with memory. There is also a *GPS* unit and *RF* transceiver (simply sketched in Figure 1) [7], [8].

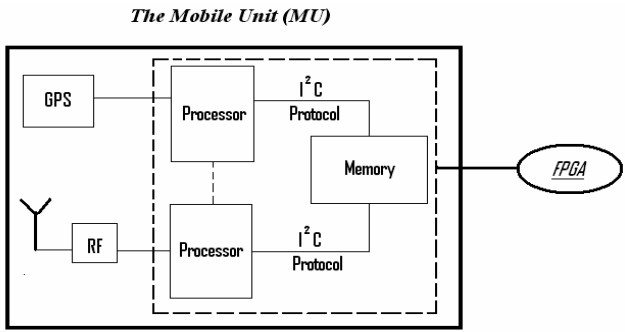

Fig.1. Modules of the *FPGA* System.

The processor of the *Aram* follows a sequential procedure of different parallel components. For instance, although two processes *P1* and *P2* hardware both exist parallel, these processes run sequentially. An interrupt, sent by the I2C controller, is used to activate one of the processes (*P1* or *P2*). But the inner part of the processes contain several parallel operations like bit assignments and selectors (corresponding to if-statements). Figures 2, 3, and 4 explain the general behavior of the different components of the *Aram*. The sequence diagram shows the sequence of messages exchanged by the set of objects performing a certain task (see Figure 2). The state diagram describes the behavior of a system, some part of a system, or an individual object (see Figure 3). The Collaboration diagram in Figure 4 emphasizes how the objects interact.

## III. The FPGA-Based Aram System

The microcontrollers make use of the same memory using a specific protocol. The system is performing properly and has a big demand in the market. However, *FPGA*s promise a better design, with a more cohesive architecture that would save processing time and speeds up system interaction. The two microcontrollers along with memory would be incorporated into or better supported with a high-density *PLD*. This will transform the hard slow interface between them into a faster and reliable programmable interconnects, and therefore makes future updates simpler. This design estimated to save a considerable percentage of the overall cost of one working unit. For a large number of demands, there would be a significant impact on production and profit. Hence, *PLD*s such as *FPGA*s are the way, for a better design in terms of upgradeability and speed, and it is a promising advancement for the production cost and revenue. The block diagram of the *FPGA* system is depicted in Figure 5.

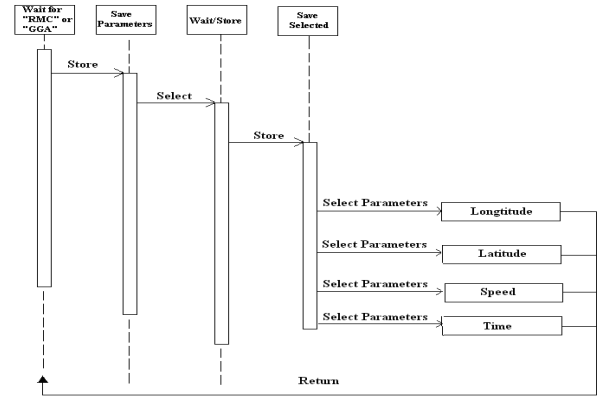

Fig.2. The sequence diagram of the *FPGA* system.

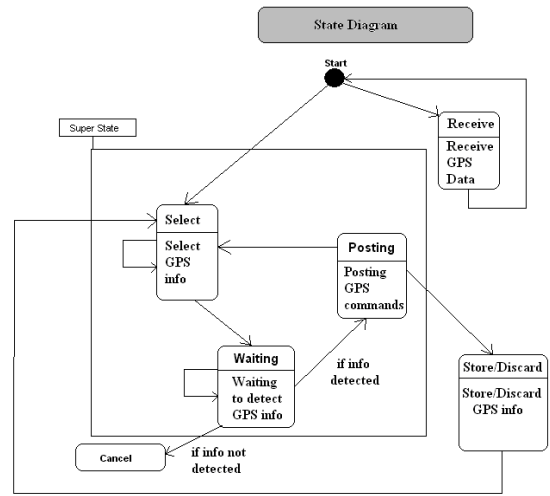

Fig.3. The state diagram of the *FPGA* system.

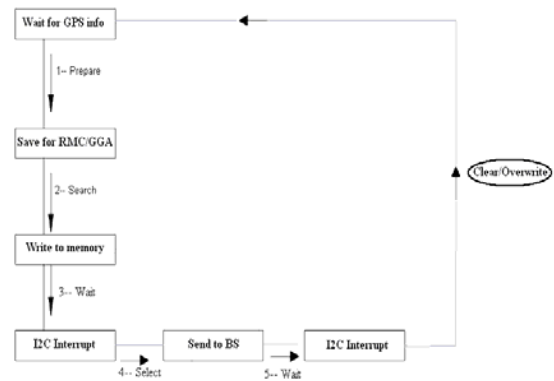

Fig.4. The collaboration diagram of the *FPGA* system.

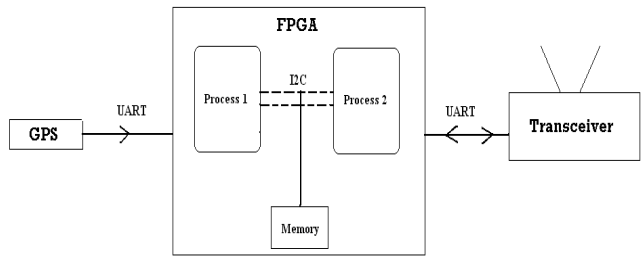

Fig.5. Block Diagram of the *FPGA* System.

Hiding the detailed architecture of the underlaying *FPGA*, The proposed system is of two communicating processes, *P1* and *P2*, along with a shared memory. In addition to the *FPGA*-based system, the *GPS* antenna and the mobile unit play significant roles. The memory block of the microcontroller-based design is replaced by hardware entity controlled by the *I2C*.

*A. Process I*

This process has to deal with the message received from the *GPS*. The default communication parameters for *NMEA* (the used protocol) output are 9600 bps baud rate, 8 data bits, stop bit, and no parity. The message includes information messages as shown in Table 1.

$GPGGA,161229.487,3723.2475,N,12158.3416,W,1,07, 1.0,9.0,M, , , ,0000*18
$GPGLL,…$GPGSA,…$GPGSV,…$GPGSV,…
$GPRMC,161229.487,A,3723.2475,N,12158.3416,W,0.13, 309.62,120598 ,*10, $GPVTG,…$GPMSS,…$GPZDA,…

From these *GPS* commands, only necessary information is selected (i.e. longitude, latitude, speed, date, and time). The data needed are found within the commands *RMC* and *GGA*; others are of minor importance to the *FPGA*. The position of the needed information is located as follows:

$GPRMC: **<time>,** <validity>, **<latitude>,** latitude hemisphere, **<longitude>,** longitude hemisphere, **<speed>,** <course over ground>, <date>, magnetic variation, check sum [7], [8], [9].

$GPGGA, **<date>,** latitude, latitude hemisphere, longitude, longitude hemisphere, <GPS quality>, <# of satellites>, horizontal dilution, <altitude>, Geoidal height, DGPS data age, Differential reference, station Identity (*ID*), and check sum. This information is stored in memory for every position traversed. Finally and when the *VTU* reaches its base station (*BS*), a large number of positions is downloaded to indicate the route covered by the vehicle during a time period and with a certain download speed. The sequential behavior of the system appears in the flow chart of Figure 6.

Initially, a flag *C* is cleared to indicate that there's no yet correct reception of data. The first state is "*Wait for GPS Parameters*", as mentioned in the flow chart, there's a continuous reception until consecutive appearance of the *ASCII* codes of "*R,M,C*" or "*GGA*" comes in the sequence. For a correct reception of data, *C* is set (ie. *C*= "*1*"), indicating a correct reception of data, and consequently make the corresponding selection of parameters and saves them in memory. When data storing ends, there is a wait state for the *I2C* interrupt to stop *P1* and start *P2*, *P2* download the saved data to the base station (*BS*). It is noted that a large number of vehicles might be in the area of coverage, and all could ask for reserving the channel with the base station; however, there are some predefined priorities that are distributed among the vehicles and therefore assures an organized way of communication. This is simply achieved by adjusting the time after which the unit sends its *ID* when it just receives the word "free".

Table I
THE PARAMETERS SENT BY THE GPS

| NMEA | Description |
|------|-------------|
| HPGGA | Global Positioning system fixed data |
| GPGLL | Geographic position-latitude/longitude |
| GPGSA | GNSS DOP and active satellites |
| GPGSV | GNSS satellites in view |
| GPRMC | Recommended minimum specific GNSS data |
| GPVTG | Course over ground and ground speed |
| GPMSS | Radio-beacon Signal-to-noise ratio, signal strength, |
| GPZDA | PPS timing message (synchronized to PPS) |

*B. Process II*

As mentioned earlier, the Base station is continuously sending the word "free", and all units within the range are waiting to receive it and acquire communication with the transceiver. If the unit received the word "free", it sends its *ID* number, otherwise it resumes waiting. It waits for acknowledge, if Acknowledge is not received, the unit sends its *ID* number and waits for feedback. If still no acknowledgement, the communication process terminates, going back to the first step. If acknowledge is received, process 2 sends Interrupt to process 1, the latter responds and stops writing to memory.

Process 2 is then capable of downloading information to the base station. When data is transmitted, the unit sends the number of points transmitted, to be compared with those received by the base station. If they didn't match, the unit repeats downloading its information all over again. Otherwise, if the unit receives successful download, it terminates the process and turns off.

Initially, the circuit shown in Figure 7 is off. After car ignition, current passes through *D1*, and continues its way towards the transistor. This causes the relay to switch and supports an output voltage of 12V. The circuit (*C\**) is now powered and could start its functionality. Using the 2 regulators, it becomes feasible to provide an adequate voltage to the *FPGA*, which in turn navigates the whole switching technique of the system. In other words, the *FPGA* adapts itself so that it can either put a zero or 5V at the side connecting D2. For the 5V, the circuit is all on, and the vehicle is in normal functionality. When data download ends, the *FPGA* perceives that, and changes the whole circuit into an idle one, and waits for another car ignition.

So, it is well known now that the *FPGA* will be the controller of the behavior of the *VTS* system.

*C. Memory*

The suggested memory blocks are addressed by a 12-bit address bus and stores 8-bit data elements. This means that the memory can store up to 4 KB of data. The memory controller navigates the proper memory addressing. Multiplexers are distributed along with the controller to make the selection of the addressed memory location and do the corresponding operation.

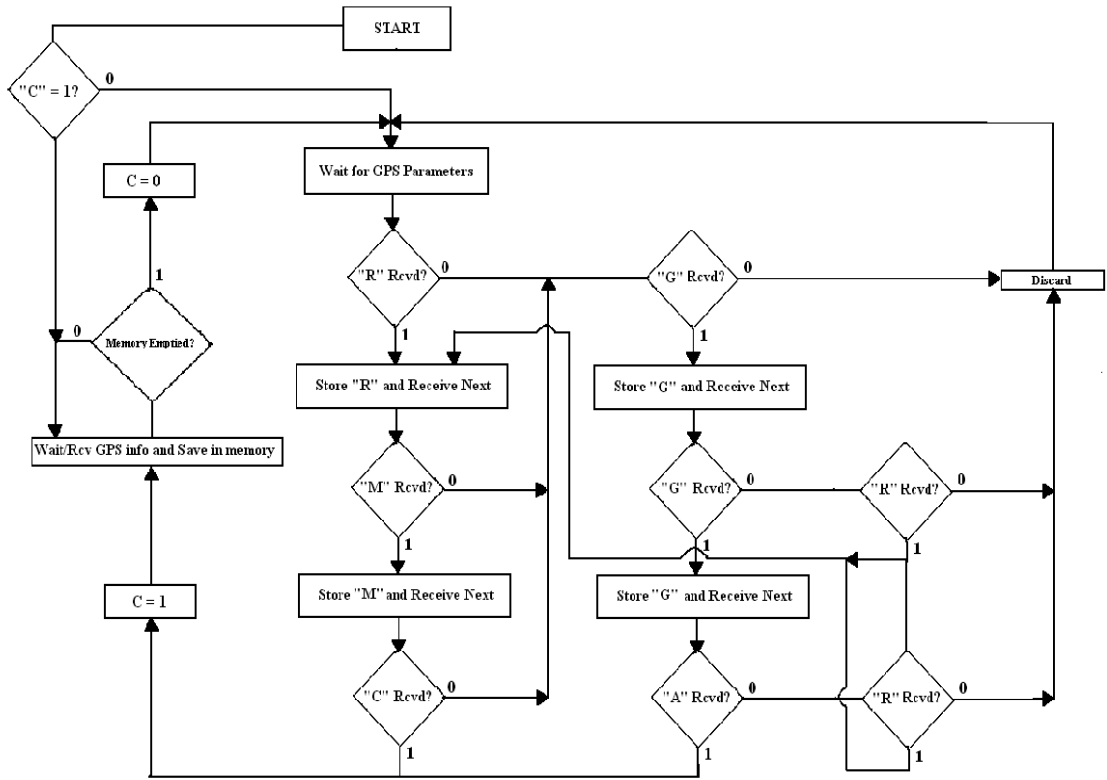

Fig. 6. The flow chart governing the main part of the *FPGA*-based System.

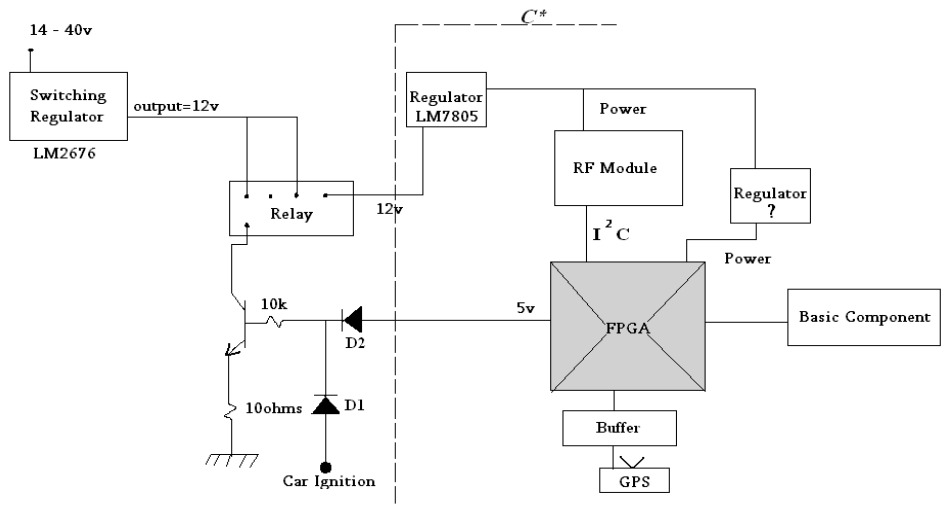

Fig.7. The electric components driving the *FPGA*.

### D. Communication Protocols: I2C and UART

The *I2C* bus is a serial, two-wire interface, popularly used in many systems because of its low overhead. It is used as the interface of process 1 and process 2 with the shared memory. It makes sure that only one process is active at a time, with good reliability in communication. Therefore, it writes data read from the *GPS* during process 1, and reads from memory to output the traversed positions into the base station. The Universal Asynchronous Receiver Transmitter (*UART*) is the most widely used serial data communication circuit ever. *UART* allows full duplex communication over serial communication links as RS232. The *UART* is used to interface *Process 1* and the GPS module from one side, and *Process 2* and the Base Station (*BS*) from the other side.

### IV. PERFORMANCE ANALYSIS AND EVALUATION

Three different systems are to be tested for the *FPGA* implementation. The suggested systems gradually add more parts to the designed FPGA implementation till we reach a complete stand alone system. The three suggested implementations are as follows:

1) *process 1*, *process 2*, and *I2C*
2) *process 1*, *process 2*, *I2C*, and memory
3) *process 1*, *process 2*, *I2C*, *UART*, and memory (standalone *FPGA*).

According to the local market cost, around %9.6 could be saved per unit if the FPGA-based all-in-one system is adopted.

For the kind of memory implemented in the system, the vehicle cannot store many locations, so the Vehicle Tracking System (*VTS*) is to be used within a small city. If the sampling rate is to be set one reading every two minutes, one could get a general but not very specific overview of the tracks traversed.

The vehicle tracking needs 344 bits of data to store the 5 important parameters (longitude, latitude, speed, date, and time). Consequently, this would need 43 memory locations. In other words, the system needs 43 locations for one reading lapsed 2 minutes.

With 4096 memory locations, the system makes 95.25 readings, meaning that the vehicle should come back to its base station every 3 hours and 10 minutes to download its information. This would be 4 hours and 45 minutes if the rate is one reading every 3 minutes. This is not very satisfactory but is tolerated as long as the intended area has a small range. However, this is one problem subject to upgradeability. One solution is to use an *FPGA* with a large memory.

Table 2 shows the results of simulation done on integration of parts (modules) forming the *FPGA*-based system. Each integrated system is tested on the component, integration, and system levels.

The 1$^{st}$ design used 1910 logic elements out of 10570 (*STRATIX EP1S10F484C5*, 175.47 MHz), and a maximum operating frequency of 8.149 *MHz*, leaving 82 % free space of the *FPGA* capacity. However, after adding memory to the integration, the number of logic elements increased to 5303, with 50% usage of the capacity. The propagation delay decreased slightly inside the *FPGA*, The decrease in propagation delay means that the optimizer found a better way to reduce its critical path.

Table II
RESULTS SUMMARY TAKEN BY THE SEYNTHESES OF INTEGRATIONS ON *STRATIX EP1S10F484C5*. THE EXECUTION TIME IN SOME CASES VARIES ACCORDING TO "RMC", "GGA", OR "FREE".

| Integ. | % Area in logic Elem. | Prop. Delay (ns) | Execution Time (ns) | Max. Op. Freq. (MHz) |
|---|---|---|---|---|
| 1 | 18% | 118.77 | Varies | 8.149 |
| 2 | 50% | 116.00 | Varies | 8.62 |
| 3 | 50% | 94.47 | Varies | 10.58 |

Similar results are shown when the *UART* part is added (standalone *FPGA)*, with an improvement in propagation delay. Although the number of logic elements has increased, it contributed to better interaction among the parts and raised the operating frequency to 10.58 MHz. Therefore, integration of parts has enhanced the delay with an expected increase in number of logic elements positively affects the processing speed when propagation finds its paths among combinations of gates. Suppose that the *GPS* message *"M"* received by the *UART* has come in the following sequence:

$GPGGA,161229.487,3723.2475,N,12158.3416,W,1,07,1.0,9.0,M, , , ,0000*18
$GPGLL,3723.2475,N,12158.3416,W,161229.487,A*2C
$GPRMC,161229.487,A,3723.2475,N,12158.3416,W,0.13,309.62,120598 ,*10

"*M*" is to be tested on the three obtained integrations (Table 3)), taking into account that the *GPS* parameters needed for the application come as represented in Section 3. The system takes selective parameters according to their position in the sequence, and checks if they correspond to the desired information. Every character is represented by its 8-bit *ASCII* representation. Table 3 shows an exact interpretation of data when received from the *GPS* and processed via several modules of the integration. After integrations have been inspected, the proposed system synthesized on different *FPGA*s, and the results appear in Table 4 [7], [8].

From the readings of Table 4, the following could be concluded testing the all-in-one system:

- *STRATIX EP1S10F484C5* (175.47 MHz) has enough number of logic elements, and the capacity taken by the project is one half its total capacity. The propagation delay is 94.474ns, thus, the system runs with a frequency of 10.58 MHz

- *STRATIX-II EP2S15F484C3* (420 MHz) has a larger number of logic elements, so the project took lesser capacity (36%), and caused more propagation delay.

- The project fits 90% in *Cyclone EP1C6F256C6* (405 MHz), but with minimum propagation delay of 90.371ns and thus 11.065MHz operating frequency.

- *APEXII EP2A15B724C7* (150 MHz) has the largest capacity among the listed devices allocating 30% only for the *ARAM* project with a largest propagation delay (181.418ns) and minimum frequency (5.512MHz).

**Table III**
**THE READINGS OBTAINED FROM THE INTEGRATIONS COMPILED ON *STRATIX EP1S10F484C5*.**

| Integ. | Size (bits) | Number of clock cycles | Prop. Delay (ns) | Speed of Processing (μs) |
|---|---|---|---|---|
| 1 | 1448 | 181 | 118.773 | 21.497 |
| 2 | 1448 | 181 | 116.003 | 20.996 |
| 3 | 1448 | 181 | 94.474 | 17.099 |

The best implementation happens when a device suits the project with maximum percentage, since it induces lower propagation delay and higher frequency. However, this does not always perfect the implementation. This depends on the real operation of the project when downloaded to the *FPG*. Also, the purpose of the design is significant to the decision, with *FPGAs* that have much memory remaining; one could use it if the purpose it to make future updates like data compression and optimization related to power dissipations.

On the other hand, processing speed is another point of discussion where memory space appears to be inversely proportional. Finally, one can setup his mind on the device best suitable for his implementation. From the readings above, it's shown that *STRATIX* family has the best compromise of capacity, propagation delay, and frequency. Moreover, it delivers better cost and performance attributes.

## V.  CONCLUSION

In this paper, we have presented an alternative design of an existing modern *GPS*-based *VTS* using *FPGAs*. The performance of the proposed implementations is evaluated on different *FPGAs*. The suggested designs show enhancement in terms of speed, functionality, and cost.. Future work includes refining the proposed designs in order to eliminate the sequential alternation of the two main internal processes, and investigating larger buffering by providing more memory elements and using state-of-art *FPGAs*.

TABLE IV
SYNTHESES OF THE VTS ON DIFFERENT FPGAs. THE EXECUTION TIME IN SOME CASES VARIES ACCORDING TO "*RMC*", "*GGA*", or "*FREE*".

| FPGA | Logic Area in Logic Elements | Prop. Delay (ns) | Exec. Time (ns) | Max. Freq. (MHz) |
|---|---|---|---|---|
| STRATIX EP1S10F484C5 | 50% | 94.474 | Varies | 10.58 |
| STRATIX-II EP2S15F484C3 | 36% | 151.296 | Varies | 6.609 |
| MAX3000A EPM3032ALC44-4 | Does not Fit | NA | NA | NA |
| Cyclone EP1C6F256C6 | 90% | 90.371 | Varies | 11.06 |
| FLEX6000 EPF6016TI144-3 | Does not Fit | NA | NA | NA |
| APEXII EP2A15B724C7 | 30% | 181.418 | Varies | 5.512 |

ACKNOWLEDGMENT

Adnan I. Yakzan thanks his parents and sister for their moral and academic support through all the academic life. Adnan stresses the valuable guidance of Dr. Issam Damaj and Dr. Rached Zantout throughout this work, in addition to the appreciation and support rendered by EDM Company and the Electrical Lab of Hariri Canadian University.